\documentclass{article}

\usepackage[final]{neurips_2024}

\usepackage[utf8]{inputenc} % allow utf-8 input
\usepackage[T1]{fontenc}    % use 8-bit T1 fonts
\usepackage{hyperref}       % hyperlinks
\usepackage{url}            % simple URL typesetting
\usepackage{booktabs}       % professional-quality tables
\usepackage{amsfonts}       % blackboard math symbols
\usepackage{nicefrac}       % compact symbols for 1/2, etc.
\usepackage{microtype}      % microtypography
\usepackage{xcolor}         % colors

\usepackage{caption}
\usepackage{subcaption}
\usepackage{microtype}
\usepackage{graphicx}
\usepackage{amsmath}
\usepackage{amssymb}
\usepackage[capitalize]{cleveref}
\usepackage{siunitx}
\usepackage{algorithm}
\usepackage{algpseudocode}
\usepackage{wrapfig}

\DeclareMathOperator*{\argmin}{arg\,min}

\title{Systems-Structure-Based Drug Design}

\author{%
  Vincent D.~Zaballa \thanks{email: vzaballa@uci.edu} \\
  Department of Biomedical Engineering\\
  University of California, Irvine\\
  Irvine, CA \\
  \And
  Elliot E. Hui \\
  Department of Biomedical Engineering\\
  University of California, Irvine\\
  Irvine, CA \\
}

\begin{document}

\maketitle

\begin{abstract}
  Recent advances in generative deep learning have transformed small molecule design, but most methods lack biological systems context, focusing narrowly on specific protein pockets. We introduce a non-differentiable diffusion guidance method that integrates systems biology models, enhancing small molecule generation with pathway context. Using the Bone Morphogenetic Protein (BMP) pathway, we generate small molecules specific to one protein over competing proteins. This method enhances the precision and efficiency of small molecule drug discovery by incorporating systems biology insights into generative models.
\end{abstract}

\section{Introduction}

Structure-based drug discovery (SBDD) is a valuable tool for discovering potent small molecule drugs \cite{anderson2003process}. SBDD operates off the principle that treating a disease can be reduced to targeting a protein relevant in the disease process, whether directly causal or simply correlated with the disease process. Using SBDD to develop new small molecule therapies provides promise for treating a wide range of diseases, such as more targeted cancer therapeutics \cite{zhong2021small}. Enzymes are an example of an excellent target class for SBDD where the active site is typically known and evolutionarily conserved, thus a drug can be created to inhibit this active site to prevent the operation of the protein involved in the disease process. Notable examples are the creation of a small molecule inhibitor of the SARS-CoV-2 (COVID-19) protease \cite{liu2022development} and HIV protease \cite{lam1994rational}. Recent deep learning methods have naturally fit into the paradigm of SBDD by generating small molecules often conditioned on some type of relevant physical information. These methods are a great fit for inhibiting proteins with known active sites.

% \citet{du2024machine} provide a comprehensive review of the various deep learning based methods for small molecule development. We focus on diffusion-based approaches for  joint generation of a small molecule for the pocket and explicitly conditioned on various forms of physical information. 

On the other hand, there are many proteins that do not have a clear active site, such as those involved in protein-protein interactions (PPIs). Binding to an active site in a PPI, known as orthosteric modulation, depends on finding ``hot spots'' via mutageneis analysis of the interface of the two proteins, and which may also depend on the plasticity of the proteins involved \cite{ran2018inhibitors, arkin2014small, cukuroglu2014hot}. While mutagenesis can reveal conserved residues at the interface, it assumes knowledge of how proteins may interact in the first place. Assuming knowledge of how two proteins interact, designing a small molecule that is specific to the PPI of interest disregards how the small molecule may interact with similar proteins, such as protein receptor complexes in the TGF-$\beta$ signaling pathway. This non-specific development likely contributes to downstream failures in clinical trials due to toxicity of the drug and related off-target effects \cite{xie2011structure}. This risk can be mitigated by further screening for off-target effects but this can be prohibitively expensive and introduces another small molecule screening step that lengthens the process of drug development \cite{dawidowski2017inhibitors}. There is a need for developing small molecules that are \textit{specific} to the target of interest, yet a paucity of machine learning methods to do this. A notable exception is work by \citet{harris2023flexible} although their method optimizes small molecules specific to one kinase over another assuming a known binding site in each kinase \textit{a priori}.

\begin{figure*}[t]
\centering
\includegraphics[trim=3.32cm 11cm 12cm 1.15cm,clip,width=\textwidth]{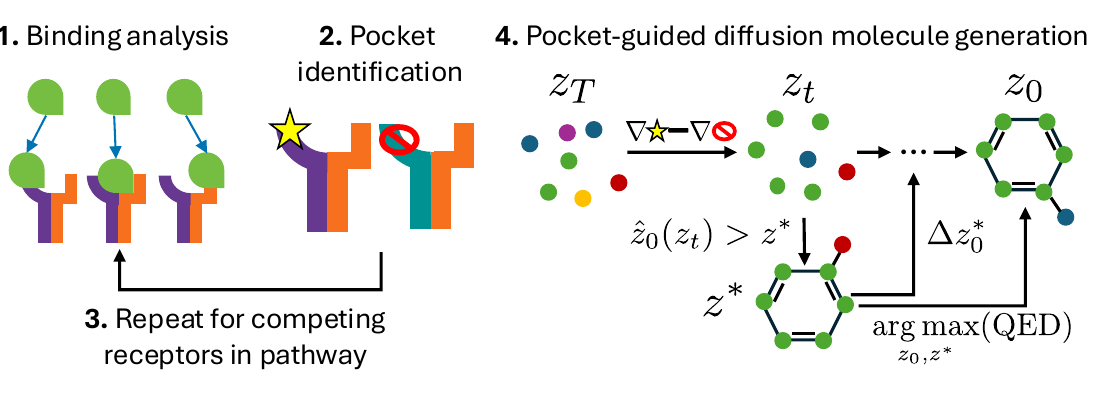}
\caption{Overview of Systems-Structure-Based Drug Discovery (SSBDD) identifying orthosteric targets and creating a small molecule specific to the target of interest. \textbf{1.} Perform a binding analysis to understand which configuration of the protein ligand and the receptor complex is most likely using systems biology data. \textbf{2.} Identify the most important point on the protein surface using mutagenesis. \textbf{3.} Repeat process for competing protein complexes in the pathway. \textbf{4.} Using known orthosteric target sites, perform \textit{forward} guided diffusion conditioning generation steps on the gradient of the desired and undesired target sites. Checkpointing is performed on denoised predictions and checkpointed molecules are used in non-differentiable \textit{reverse} guidance. At the end of sampling, the checkpointed molecule is compared according to a metric with the final denoised molecule and the best is returned. }
\label{fig:struct+systems}
\vskip -12pt
\end{figure*}

We demonstrate a bioinformatics pipeline and machine learning method to generate small molecules that are more specific to a protein in a signaling pathway involved in PPIs. The bioninformatics process helps select which proteins may be of interest to target and at which sites to target. The machine learning method is based on an a non-differentiable guided diffusion sampling process conditioned on how well small molecules dock specifically to a target of interest in the pathway of interest.  Finally, we evaluate the combined pipeline in an example of the Bone Morphogenetic Protein (BMP) pathway \cite{wang2014bone}, showing how we are able to guide small molecule generation to inhibit one protein's PPI active site while minimizing the likelihood of binding to another protein in the pathway that shares a protein ligand in the pathway. Given the inclusion of systems biology information \cite{alon2019introduction}, we introduce Systems-Structure-Based Drug Design (SSBDD).

\section{Background}
\label{background}

\textbf{Diffusion Models} Diffusion models are a class of deep learning method capable of learning the gradient, or score, of a denoising process going from a noisy distribution $\boldsymbol{z}_T$ to the data distribution $\boldsymbol{z}_0$ by a series of diffusion steps \cite{song2020generativemodelingestimatinggradients, ho2020denoisingdiffusionprobabilisticmodels}. Diffusion has been used in a wide variety of biological tasks such as docking \cite{corso2022diffdock}, protein generation \cite{ingraham2022illuminating, watson2023novo}, modeling protein conformation landscapes \cite{lu2024dynamicbind}, and small molecule generation \cite{schneuing2022structure, hoogeboom2022equivariantdiffusionmoleculegeneration, morehead2024geometry}.

We follow the notation of \cite{hoogeboom2022equivariantdiffusionmoleculegeneration} in describing our diffusion process. Data samples are point clouds $\boldsymbol{z}_0 = [\boldsymbol{x}, \boldsymbol{h}]$ which represent the 3D coordinates of $N$ atoms $\boldsymbol{x} \in \mathbb{R}^{N \times 3}$ and atom features $\boldsymbol{h} \in \mathbb{R}^{N \times d}$. Given a data point $\boldsymbol{x}$, a step in the diffusion process is defined as:
\begin{equation}
    q(\boldsymbol{z}_t | \boldsymbol{x}) = \mathcal{N} (\boldsymbol{z}_t | \alpha_t \boldsymbol{x}_t, \sigma^2_t \mathbf{I}),
    \label{noise_kernel}
\end{equation}
where $\alpha_t \in \mathbb{R}^+$ controls signal retention and $\sigma_t \in \mathbb{R}^+$ controls the addition of noise. Typically $\alpha_t$ is modeled by a function that smoothly transitions from $\alpha_0 \approx 1$ to $\alpha_T \approx 0$. The noising process can be described by a variance preserving process \cite{sohl2015deep, ho2020denoisingdiffusionprobabilisticmodels} where $\alpha_t = \sqrt{1 - \sigma_t^2}$. We can also define the signal to noise ratio $\text{SNR}(t) = \alpha_t^2 / \sigma_t^2$ \cite{kingma2021variational} to simplify the notation. Since the diffusion is Markovian, we can can write the transition from a time step $t$ to $s < t$ as
\begin{equation}
    q(\boldsymbol{z}_s \mid \boldsymbol{z}_{\text{data}}, \boldsymbol{z}_t) 
    = \mathcal{N} \left( \boldsymbol{z}_s \big| 
    \frac{\alpha_{t|s} \sigma_s^2}{\sigma_t^2} \boldsymbol{z}_t 
    + \frac{\alpha_s \sigma_{t|s}^2}{\sigma_t^2} \boldsymbol{z}_{0}, 
    \frac{\sigma_{t|s}^2 \sigma_s^2}{\sigma_t^2} \mathbf{I} \right),
\end{equation}
where $\alpha_{t|s} = \alpha_t / \alpha_s$ and $\sigma_{t|s}^2 = \sigma_t^2 - \alpha_{t|s}^2 \sigma_s^2$ following notation of \cite{hoogeboom2022equivariantdiffusionmoleculegeneration}. The true denoising process depends on the data $\boldsymbol{z}_0$ which is unavailable when generating new samples. However, an approximate sample $\hat{\boldsymbol{z}}_0$ is calculated using a neural network $\phi_\theta$. Thus, \cref{noise_kernel} can be reparameterized as $\boldsymbol{z}_t = \alpha_t \boldsymbol{z}_0 + \sigma_t \boldsymbol{\epsilon}$ with $\epsilon \sim \mathcal{N}(\boldsymbol{0}, \boldsymbol{I})$ to directly predict the Gaussian noise $\hat{\boldsymbol{\epsilon}}_\theta = \phi_\theta (\boldsymbol{z}_t ,t)$. We can rearrange to show the denoised prediction as a function of the current noisy prediction $\hat{\boldsymbol{z}}_0 ( \boldsymbol{z}_t ) = \boldsymbol{z}_t / \alpha_t - \hat{\boldsymbol{\epsilon}}_\theta \sigma_t / \alpha_t$. We diverge from other notation by showing the prediction as a function of the noised sample at time $t$. The neural network is trained by maximizing the variational lower bound of the data distribution, equivalent to the simplified training objective $\mathcal{L}_\text{train} = \frac{1}{2} || \hat{\boldsymbol{\epsilon}} - \phi_\theta(\boldsymbol{z}_t, t) ||^2$ up to a scaling factor \cite{ho2020denoisingdiffusionprobabilisticmodels, kingma2021variational}.

\textbf{Diffusion Guidance}   There has been great interest in guiding diffusion models for image generation that range from adding explicit conditional information during the training process \cite{ho2022classifierfreediffusionguidance, nichol2022glidephotorealisticimagegeneration} and guiding the diffusion generation process using a separate and differentiable guidance function after training \cite{dhariwal2021diffusionmodelsbeatgans, wang2022zeroshotimagerestorationusing}. \citet{bansal2023universal} demonstrated a general-purpose method for guided diffusion simply using a differentiable guidance function. They split guidance into forward and backward guidance. Forward guidance during sample generation updates the noise prediction using a weighted gradient of the guidance function
\begin{equation}
    \hat{\boldsymbol{\epsilon}}_\theta (\boldsymbol{z}_t, t) = \boldsymbol{\epsilon}_\theta (\boldsymbol{z}_t, t) + s(t) \cdot \nabla_{\boldsymbol{z}_t} \ell (c, f(\hat{\boldsymbol{z}}_0 (\boldsymbol{z}_t)))
    \label{forward_guidance}
\end{equation}
where we can use the prediction of the denoised data point at time $t$ as $\hat{\boldsymbol{z}}_0(\boldsymbol{z}_t)$, $s(t)$ is a hyperparameter for the  guidance strength, $c$ is contextual information such as a protein pocket, and $\ell(\cdot, \cdot)$ is a function that depends on the context and denoised prediction. Backward guidance helps to ensure the generated data point satisfies the guidance function. Instead of calculating the gradient w.r.t. the denoised prediction, we calculate the difference in clean data space $\Delta \boldsymbol{z}_0$ by optimizing
\begin{equation}
    \Delta \boldsymbol{z}_0 = \argmin_{\Delta} \ell (c, f(\hat{\boldsymbol{z}}_0 + \Delta))
\end{equation}
where the $\Delta$ can be optimized using $m$-step gradient descent starting at $\Delta = 0$. Once we find the $\Delta$ that optimizes the guidance function, we can translate the change back to the original noised data space and calculate the guided denoising prediction $\hat{\boldsymbol{\epsilon}}$ as $\boldsymbol{z}_t = \sqrt{\alpha_t} (\hat{\boldsymbol{z}}_0 + \Delta \boldsymbol{z}_0 ) + \sqrt{1 - \alpha_t} \hat{\boldsymbol{\epsilon}}$. Using the definition of the predicted denoised data point from the current noised data point, the augmentation to the original denoising prediction is
\begin{equation}
    \hat{\boldsymbol{\epsilon}} = \boldsymbol{\epsilon}_\theta (\boldsymbol{z}_t, t) - \sqrt{\alpha_t / 1 - \alpha_t} \Delta \boldsymbol{z}_0.
    \label{reverse_guidance}
\end{equation}

\textbf{Equivariant Graph Neural Networks}  Atoms live in a space that is invariant to translations and rotations. To build these inductive biases into the neural network $\phi_\theta$, we use a Graph Neural Network (GNN) that is a function $f: \mathcal{X} \rightarrow \mathcal{Y}$ equivariant with respect to the group $G$ if the action of the group element, $g \in G$, on the function is equivalent to the input, i.e. $f(g.\boldsymbol{x}) = g. f(\boldsymbol{x})$ where $g.$ represents the group action. We follow \cite{morehead2024geometry} and use the \textsc{GCPN}\scalebox{0.8}{ET}\textsc{++} GNN model that maintains SE(3) equivariance, which is equivariant to 3D roto-translations, while maintaining chirality of molecules by using Geometry-Complete Perceptron Networks (\textsc{GCPN}\scalebox{0.8}{ETS}) that define direction-robust local geometric reference frames \cite{morehead2023geometrycompleteperceptronnetworks3d}.

\textbf{Molecular Docking}  Docking of small molecule candidates is split between known-pocket and blind docking. \cite{corso2022diffdock} demonstrated the use of diffusion methods for the blind docking setting while achieving state of the art results at the time. Related work for known-pocket diffusion, DiffDock-Pocket \cite{plainer2023diffdock}, predicts ligand poses in a specific binding pocket of interest, such as orthosteric sites, by modeling the flexibility of the side chains of a binding pocket that would influence the conformation of the docked small molecule. Recent methods have focused on the problem of adequately modeling the protein conformation when unbound, its apoprotein state, and when bound to a molecule, its holoprotein state \cite{corso24flexible, lu2024dynamicbind}. Regardless of the scope, most diffusion-based docking methods include a confidence model that provides a score of how confident the model is in its prediction of a ligand to the binding spot. 

% Given our setting of focusing on orthosteric binding pockets we use the DiffDock-Pocket diffusion model to model the flexibility of side chains when generating a small molecule. 

\textbf{Systems Biology \& Protein-Protein Interactions}  Understanding how proteins interact is critical to develop therapies for PPI inhibitors. Systems biology is the study of biological networks, including networks of proteins. Protein network response can be modeled using mass action kinetics \cite{antebi2017combinatorial} and have physically-meaningful parameters that can be linked to protein structure predictions \cite{zaballa2024reducing}. This link provides a way to evaluate which structural interactions are likely while cross-validating with a gene reporter assay data fit to systems biology models.

% just make this a quick paragraph of the ways to do small molecule generation

\section{Systems-Structure-Based Drug Discovery}
\label{method}

\textbf{Identifying Targets}  We follow the method of \citet{zaballa2024reducing} and identify PPI candidates by evaluating structural predictions' binding affinities with systems biology data. Briefly, we perform evolutionary analysis to identify conserved regions of the proteins, dock using Ambiguous Interaction Restraints informed by conserved residues, and cross-validate a structural binding affinity prediction with systems biology data. Furthermore, we use the conserved regions at the interface of two proteins in the pathway to identify a candidate orthosteric binding pocket that can be used to generate novel molecules. We repeat this process for competing proteins in the pathway that share a diffusible protein ligand in order to find the pocket that we want to avoid hitting. More details can be found in \cref{bmp_select}.

\textbf{Non-Differentiable Diffusion Guidance}  To guide molecule generation specific to one orthosteric site and not to another, we use DiffDock-Pocket \cite{plainer2023diffdock} as the guidance function and use the confidence prediction of a generated small molecule as the guidance signal. Thus, we would like higher confidence for the pocket of interest and lower confidence for the competing protein. While the confidence model can provide a signal it is not a differentiable function, which is required for both forward and backward guidance. We overcome this by approximating the gradient of the forward diffusion guidance using the central difference form of finite difference with an offset of $\varepsilon = 1e^{-2}$ as
\begin{equation}
    \nabla_{\boldsymbol{z}_t} \ell (c, f(\hat{\boldsymbol{z}}_0)) \approx \frac{\ell (c, f(\hat{\boldsymbol{z}}_0(\boldsymbol{z}_t + \varepsilon)) - \ell (c, f(\hat{\boldsymbol{z}}_0 ( \boldsymbol{z}_t - \varepsilon)))}{2\varepsilon}.
\end{equation}

In our approach to reverse guidance, instead of perturbing the clean data point within the guidance function, we introduce a checkpointing mechanism that selects a molecule based on its guidance score as the basis for future reverse guidance steps. Specifically, at each reverse step, we evaluate the denoised prediction $ \hat{\boldsymbol{z}}_0 $ for the current $ \boldsymbol{z}_t $ and compare it to a checkpointed molecule $ \boldsymbol{z}^* $, which represents the molecule that has achieved the highest guidance score in previous iterations. The guidance score is evaluated using the quantitative estimate of drug-likeness drug-likeness metric (QED) \cite{bickerton2012quantifying}, where a higher score corresponds to a more desirable molecule. If the current denoised prediction $ \hat{\boldsymbol{z}}_0 $ achieves a higher QED score than $ \boldsymbol{z}^* $, we update $ \boldsymbol{z}^* $ to this new molecule. We then calculate an adjusted reverse diffusion gradient point as $\Delta \boldsymbol{z}_0^* = \boldsymbol{z}^*_0 - \hat{\boldsymbol{z}}_0(\boldsymbol{z}_t)$ This mechanism allows us to use the best-performing molecule as the reference for subsequent reverse guidance steps, ensuring that the generated molecule is progressively optimized according to the guidance metric. Thus we adjust \cref{reverse_guidance} to be
\begin{equation}
    \hat{\boldsymbol{\epsilon}} = \boldsymbol{\epsilon}_\theta (\boldsymbol{z}_t, t) - \sqrt{\alpha_t / 1 - \alpha_t} \Delta \boldsymbol{z}_0^*.
\end{equation}
Since this method relies on a good checkpoint, we predict a batch of samples $\hat{\boldsymbol{z}}_0(\boldsymbol{z}_t)$ and checkpoint on the best sample. Curiously, we found that this crude reverse guidance was required for convergence of non-differentiable guided diffusion. This may be due to forward guidance steering sampled molecules out of distribution of a lower-dimensional manifold and the reverse guidance providing a signal that steers the sample back to the data manifold. Finally, we can compare this checkpointed molecule to the final diffusion guidance molecule and return the molecule that achieves the best QED score. We chose checkpointing on the QED score instead of an aggregate of the confidences and the QED as it helped to stabilize the sampling process. The complete algorithm can be seen in \cref{the_algo}.

\section{Experiments}
\label{experiments}

We evaluated SSBDD on a model of the BMP pathway \cite{su2022ligand} using previously-collected data \cite{klumpe2022context} to perform the systems biology validation. In particular, we chose to design a small molecule specific to the BMPR1A-ACVR2A receptor complex in combination with the BMP4 diffusible ligand due to the high binding affinity of the complex. We chose the ACVR1-ACVR2A receptor complex due to its predicted high binding affinity to the BMP4 protein ligand. Since both complexes bind strongly to BMP4, we would like to design a small molecule that is more specific to BMPR1A-ACVR2A-BMP4 (BA4) than ACVR1-ACVR2A-BMP4 (AA4).  We found ``hot spots'' that are likely important to the PPI between each receptor complex and the BMP4 diffusible ligand using evolutionary analysis data. We used the coordinates of the ``hot spots'' as the center of the DiffDock-Pocket diffusion process. Selection of the protein complexes and evolutionary analysis is discussed in \cref{bmp_select}.

To perform the guided diffusion process, we generated 60, 75, and 90 atoms. We return the final or checkpointed molecules and then evaluated according to its QED, synthetic accessibility (SA) score, and DiffDock-Pocket confidence score in each pocket for both the pocket of interest and competing pocket. We show a list of the sampling hyperparameters in \cref{hyperparams}.

\begin{table}[t]
\caption{Metrics of small molecules generated with 60, 75, and 90 atoms using three samples each ($n=3$) on QED score, SA score, and the confidence for both the BA4 and AA4 protein complex pockets of interest. Values are reported as mean $\pm$ standard deviation. }
\vskip 3pt
\centering
\begin{tabular}{cccccc}
\hline
Number of Atoms & QED ($\uparrow$) & SA ($\downarrow$) & BA4 Conf. ($\uparrow$) & AA4 Conf. ($\downarrow$) \\ % & Vina Score ($\downarrow$) \\
\hline
60 & $\bold{0.764 \pm 0.016}$ & $6.805 \pm 1.205$ & $-0.732 \pm 0.363$ & $\bold{-7.851 \pm 0.702}$ \\ % & $\mu_1 \pm \sigma_1$ \\
75 & $0.703 \pm 0.019$ & $\bold{5.181 \pm 1.088}$ & $\bold{-0.482 \pm 1.037}$ & $-7.782 \pm 0.387$ \\ % & $\mu_2 \pm \sigma_2$ \\
90 & $0.740 \pm 0.065$ & $5.533 \pm 1.353$ & $-1.641 \pm 1.390$ & $-6.480 \pm 1.390$ \\ % & $\mu_3 \pm \sigma_3$ \\
\hline
\end{tabular}
\label{tab:molecular_metrics}
\vskip -15pt
\end{table}

\begin{wrapfigure}{r}{0.5\textwidth}
\vskip -10pt
\includegraphics[width=\linewidth]{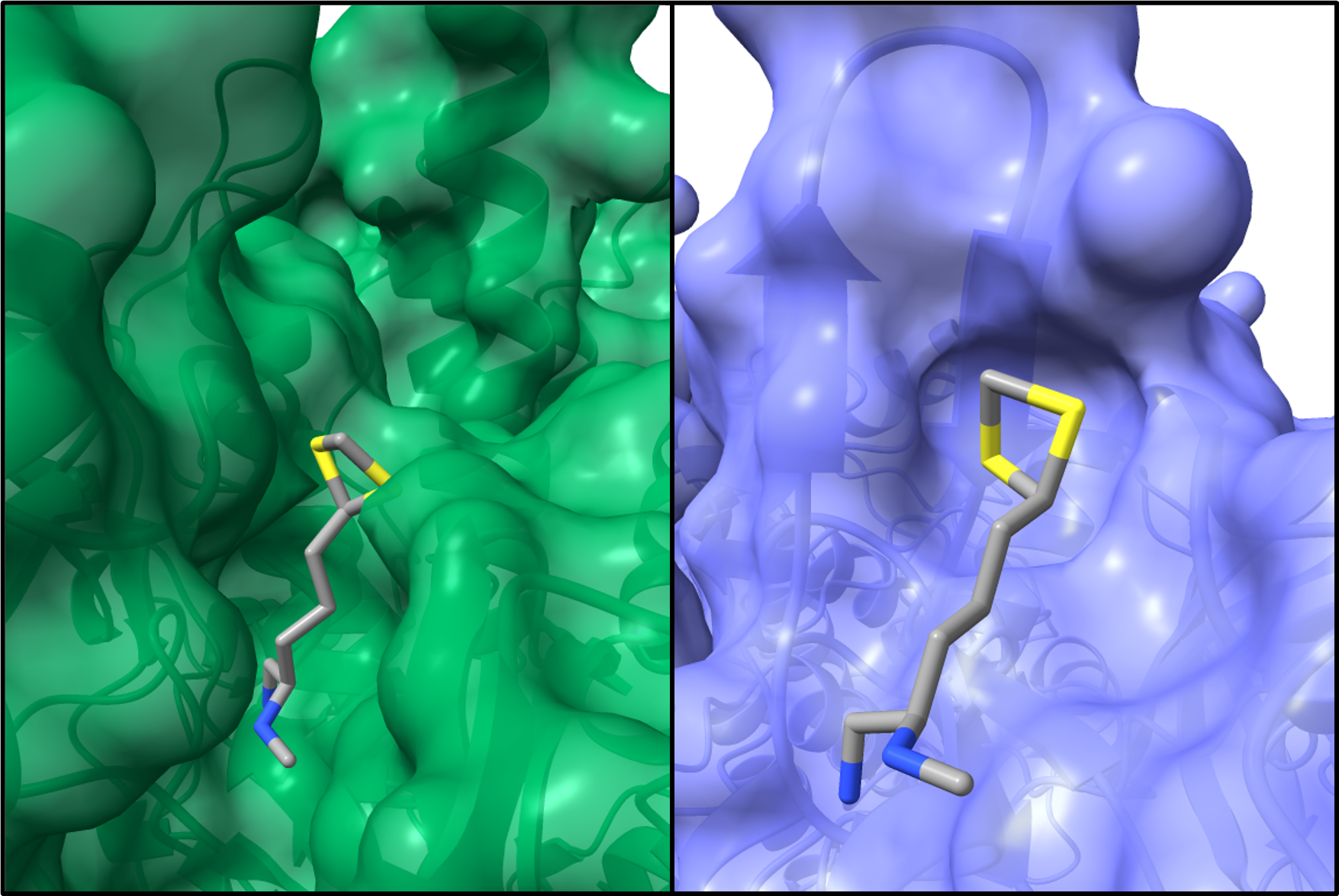}
\caption{A 75-atom small molecule generated and fit into the competing (left) and targeted (right) pocket of interest. Vina score (lower is better) for the competing complex was $\bold{4.27}$ kcal/mol and $\bold{-1.91}$ kcal/mol for the targeted pocket. }
\label{fig:pocket_compare}
\vskip -10pt
\end{wrapfigure}

\cref{tab:molecular_metrics} shows SSBDD is able to consistently generate small molecules specific to the target of interest, as demonstrated by the BA4 and AA4 confidence scores. We found that there is likely an optimal sized small molecule that specifically binds a target of interest. We see the smaller and larger molecules generally return a better QED score than the 75-atom small molecule but the 75-atom molecule is generally more confident in the target of interest than the small and large small molecules. In general, we also found that checkpointing on earlier predicted molecules generally led to smaller final molecules since we chose to keep the largest fragment of a structure prediction. This variability in size introduces more randomness in the process and treats the atom size as an upper bound of the molecule size, and suggests optimizing for the size of a small molecule is an important variable when making target-specific molecules.

We evaluated the Autodock-Vina score in one of the 75-atom small molecules in the target and competing protein pockets \cref{fig:pocket_compare} and found dramatically stronger binding affinity of the molecule to the target of interest as opposed to the competing target. This seems counterintuitive as the competing protein pocket is larger and secluded from water, but the ligand's size may prevent it from fully engaging in key bending residues. By contrast, the ligand in the targeted pocket better fits the pocket's shape likely resulting in more energetically favorable interactions, such as hydrogen bonding and van der Waals contacts, leading to a stronger overall binding affinity. 

\section{Discussion}
\label{discuss}
We demonstrated a bioinformatics pipeline for selecting a protein pocket of interest, as well as competing pockets, and a novel non-differentiable diffusion guidance method that is able to generate molecules with greater specificity to the target of interest. The non-differentiable guidance function provides greater flexibility in terms of which functions can guide the diffusion sampling process. In evaluating our method across various molecule sizes, we observed that medium-sized molecules typically exhibited superior performance in the targeted pocket; however, the optimal molecule size likely varies depending on the characteristics of both the target and competing pockets.

\textbf{Future Work}  While our method helps overcome challenges associated with non-differentiable functions, there is a penalty in sampling efficiency by having to evaluate non-differentiable functions with a batch of molecules taking 1-2 hours to sample. Creating a differentiable alternative to DiffDock-Pocket \cite{plainer2023diffdock}, e.g. using flow matching, would greatly accelerate sampling. Additionally, adapting diffusion models that generate small molecules of variable size \cite{campbell2023transdimensionalgenerativemodelingjump} would allow for a more in-depth study of the interdependence between size of the molecule and shape of binding pockets of interest.

% \section*{References}

\bibliography{main}
\bibliographystyle{neurips_2024}

%%%%%%%%%%%%%%%%%%%%%%%%%%%%%%%%%%%%%%%%%%%%%%%%%%%%%%%%%%%%

\newpage
\appendix

\section{Appendix}
\subsection{Non-Differentiable Diffusion Guidance Algorithm}
\label{the_algo}

We follow the algorithm notation from \cite{bansal2023universal} in \cref{algo} and denote the prediction of a new data point in sampling as $S(\boldsymbol{z}_t, \hat{\boldsymbol{\epsilon}}_\theta, t)$, where the subsequent diffusion step is dependent on the current noised data point, prediction, and time point. 

\begin{algorithm}
\caption{Non-Differentiable Universal Guidance}
\textbf{Parameter:} Recurrent steps $k$, number of checkpoint samples $m$, and guidance strength $s(t)$\\
\textbf{Required:} $\boldsymbol{z}_T$ sampled from $\mathcal{N}(\boldsymbol{0}, \mathbf{I})$, diffusion model $\boldsymbol{\epsilon}_\theta$, noise scales $\{\alpha_t\}_{t=1}^{T}$, guidance function $f$, loss function $\ell$, and prompt $c$
\begin{algorithmic}
\State $\boldsymbol{z}^* \leftarrow \boldsymbol{0}$
\For{$t = T, T - 1, \dots, 1$}
    \For{$n = 1, 2, \dots, k$}
        \State Calculate $\hat{\boldsymbol{z}}_0 ( \boldsymbol{z}_t ) = \boldsymbol{z}_t / \alpha_t - \boldsymbol{\epsilon}_\theta \sigma_t / \alpha_t$
        \State Calculate $\hat{\boldsymbol{\epsilon}}_\theta$ using forward universal guidance as in \cref{forward_guidance}
        \State  Sample $m$ values of $\boldsymbol{z}_{t}^{(i)} \leftarrow S(\boldsymbol{z}_{t+1}, \hat{\boldsymbol{\epsilon}}_\theta, t+1)$ for $i=1, 2, ..., m$
        \For {each sample $\boldsymbol{z}_{t}^{(i)}$, where $i = 1$ to $m$}
            \If {$\hat{\boldsymbol{z}}_0(\boldsymbol{z}_t^{(i)}) > \boldsymbol{z}^*$}
                \State Update $\boldsymbol{z}^* \leftarrow \hat{\boldsymbol{z}}_0(\boldsymbol{z}_t^{(i)})$
            \EndIf
        \EndFor
        \State $\Delta \boldsymbol{z}_0^* \leftarrow \boldsymbol{z}^* - \hat{\boldsymbol{z}}_0(\boldsymbol{z}_t )$ 
        \State Perform reverse guidance $ \hat{\boldsymbol{\epsilon}} \leftarrow \hat{\boldsymbol{\epsilon}} - \sqrt{\alpha_t / (1 - \alpha_t)} \Delta \boldsymbol{z}_0^* $
    \EndFor
    \State $\boldsymbol{z}_{t-1} \leftarrow S(\boldsymbol{z}_t, \hat{\boldsymbol{\epsilon}}_\theta, t)$
    \State $\boldsymbol{\epsilon}' \sim \mathcal{N}(\boldsymbol{0}, \mathbf{I})$
    \State $\boldsymbol{z}_t \leftarrow \sqrt{\alpha_t} \boldsymbol{z}_{t-1} + \sqrt{1 - \alpha_t} \boldsymbol{\epsilon}'$
\EndFor
\end{algorithmic}
\label{algo}
\end{algorithm}

\subsection{Sampling Hyperparameters \& Details}
\label{hyperparams}

The sampling hyperparameters used can be seen in \cref{tab:hyperparams}. Since predicting a denoised data point is difficult early in the sampling process, we do not begin \cref{algo} until half way through generation, when $t=0.5$, assuming the noise distribution is at $T=1$ and the data distribution is at $t=0$. Additionally, many time steps are unable to predict a denoised molecule so we implement the algorithm using a try/except loop, falling back to normal diffusion sampling if predicting a denoised molecule is not possible.

\begin{table}[h]
% \vspace{-5pt}
\centering
\setlength{\tabcolsep}{4pt}
\renewcommand{\arraystretch}{0.9}
\caption{Sampling hyperparameters.}
\begin{tabular}{lcccc}
        \toprule    
        Recurrent steps, $k$                      & $3$ \\
        Guidance strength, $s(t)$          &  $0.1 * \sqrt{1 - \alpha_t}$ \\
        Finite difference, $\varepsilon$              & $1e^{-2}$ \\
        Samples $m$ from predicted batch $\boldsymbol{z}_{t} \leftarrow S(\boldsymbol{z}_{t+1}, \hat{\boldsymbol{\epsilon}}_\theta, t+1)$              & $10$ \\
    % \bottomrule
\end{tabular}
% \vspace{-10pt}
\label{tab:hyperparams}
\end{table}

\subsection{Selection of BMP Signaling Proteins}
\label{bmp_select}

\begin{figure*}[th]
\centering
\includegraphics[trim=4cm 5cm 7.25cm 5cm,clip,width=\textwidth]{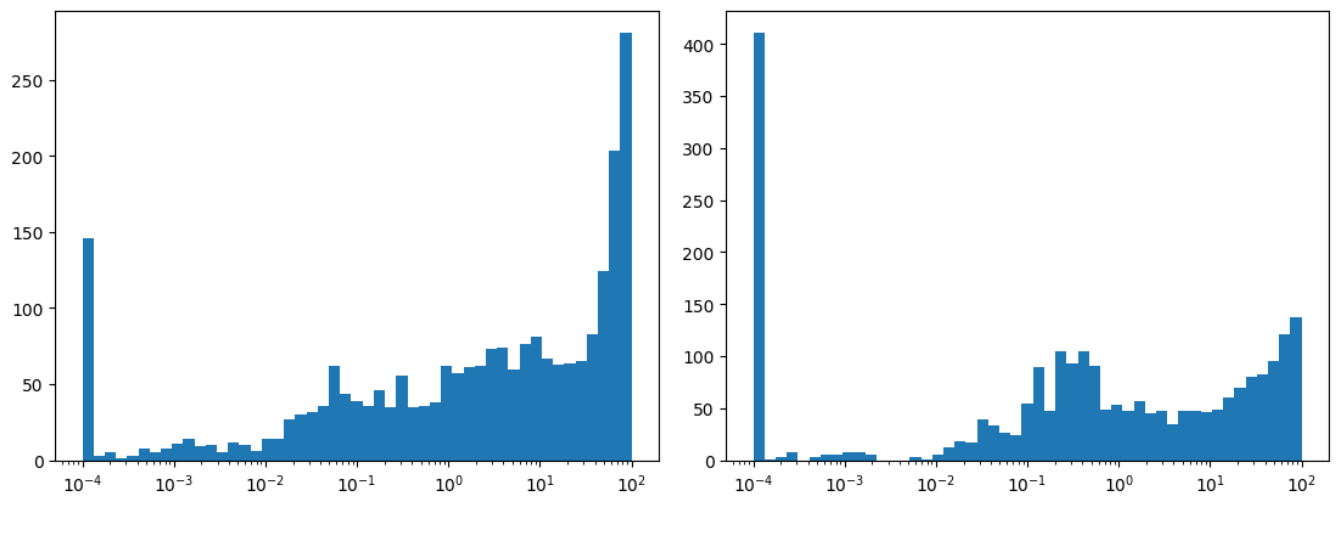}
\caption{Histogram of the posterior parameter fit of the BMP model for the ACVR1-ACVR2A-BMP4 (left) and BMPR1A-ACVR2A-BMP4 (right) trimeric complexes. Binding affinity $K_{eq}$, which is $K_{eq} = - \ln{k_d}$, is shown on the x-axis and where higher values mean more tightly binding complexes. Both suggest a strong binding affinity between the BMP4 homodimer and receptor complex.}
\label{fig:histos}
\vskip -12pt
\end{figure*}

We used a fit model of the BMP pathway to assess which complexes might be good candidates for inhibition using a small molecule as shown in \cref{fig:histos}. Given that both histograms place more density at the higher end of binding affinities, which represents stronger binding between the complex and the BMP4 homodimer, we determined they were good candidates to perform PPI inhibition. 

\begin{figure*}[h]
\centering
\includegraphics[trim=4cm 7cm 11.25cm 5.45cm,clip,width=\textwidth]{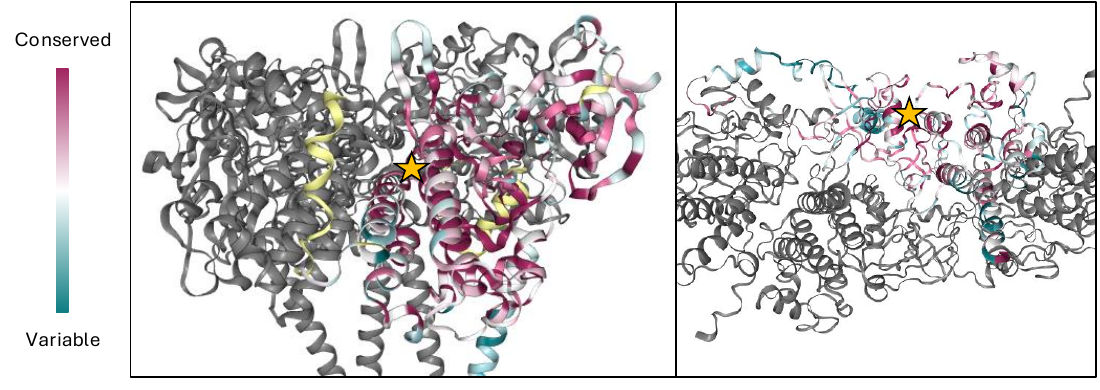}
\caption{Evolutionary analysis demonstrating ``hot spot'' in both the BMPR1A-ACVR2A (left) and the ACVR1-ACVR2A (right) receptors. The yellow start denotes a conserved point that also overlaps with a probable binding domain between each receptor complex and the BMP4 homodimer ligand. The BMPR1A-ACVR2A is shown from the side and the ACVR1-ACVR2A is shown from the top. }
\label{fig:hot_pocket}
\vskip -12pt
\end{figure*}

After determining that the two complexes were good candidates, we performed a docking analysis to determine which parts of the proteins bound with one another. We used HADDOCK \cite{van2016haddock2} and \cite{ketata2023diffdock} to find probable binding sites betwee the receptor complex and the BMP4 homodimer ligand. Additionally, after analyzing the BMP pathway and finding good PPI candidates, we performed an evolutionary analysis using the ConSurf server \cite{celniker2013consurf} to find small molecule binding ``hot spots''.  The two points can be seen in \cref{fig:hot_pocket}.

% Plot of the protein-protein interactions.

% \subsection{Sampling Curves}
% \label{sampling}

%%%%%%%%%%%%%%%%%%%%%%%%%%%%%%%%%%%%%%%%%%%%%%%%%%%%%%%%%%%%

\end{document}